\documentclass[pra,preprint,amsmath,amssymb,superscriptaddress]{revtex4-1}
\usepackage{graphicx}
\usepackage[caption=false]{subfig}
\usepackage{dcolumn}
\usepackage{bm}
\usepackage[utf8]{inputenc}
\usepackage[T1]{fontenc}
\setcitestyle{square}
\usepackage{color}
\usepackage{xcolor}
\usepackage{siunitx}
\usepackage{physics}

\DeclareSIUnit{\mrad}{\milli\radian}

\newif\ifshowcorrections
\showcorrectionsfalse
\newcommand{\QUB}{Centre for Light-Matter Interactions,
  School of Mathematics and Physics,
  Queen's University Belfast,
  BT7 1NN, Belfast, United Kingdom}

\newcommand{\Gothenburg}{Department of Physics,
  University of Gothenburg, 
  SE-41296 Gothenburg, 
  Sweden}

\newcommand{\BO}{INFN and University of Bologna, Bologna, Italy}

\newcommand{\PA}{INFN and University of Padova, Padova, Italy}

\usepackage{hyperref}       
\usepackage[capitalise]{cleveref}       
\usepackage{url}            
\usepackage{booktabs}       
\usepackage{amsfonts}       
\usepackage{nicefrac}       
\usepackage{lipsum}
\usepackage{wasysym}
\usepackage{lineno}

\usepackage{natbib}
\usepackage{float}

\begin{document}

\title{Dependence on laser intensity of the number-weighted angular distribution of Compton-scattered photon beams}

\author{K.~Fleck}
\affiliation{\QUB}

\author{T.~Blackburn}
\affiliation{\Gothenburg}

\author{E.~Gerstmayr}
\affiliation{\QUB}

\author{M.~Bruschi}
\affiliation{\BO}

\author{P.~Grutta}
\affiliation{\PA}

\author{M.~Morandin}
\affiliation{\PA}

\author{G.~Sarri}
\affiliation{\QUB}

\begin{abstract}
Inverse Compton scattering of ultra-relativistic electron beams in the field of a high-intensity laser produces photon beams with angular and spectral distributions that are strongly dependent on the laser intensity. 
Here, we show that the laser intensity at the interaction point can be accurately inferred from the measurement of the angular number-density distribution of Compton-scattered photon beams.
The theoretical expressions, supported by numerical simulations, are accurate to within 10-15\% in a wide range of laser intensities (dimensionless intensity $5 \leq a_0 \leq 50$) and electron energies (250 \unit{\MeV} $\leq E \leq $ 15 \unit{\GeV}), and accounts for experimental features such as the finite transverse size of the electron beam, low-energy cut-offs in the photon detector, and the possibility of a transverse misalignment between the electron beam and the laser focus. 
\end{abstract}


\maketitle

\section{Introduction}\label{sec:Introduction}
At the frontier of ultra-high electromagnetic intensities, it is now possible to access peak laser intensities of up to $\sim 10^{23}$ \unit{W/cm^2} \cite{Yoon:2021}, with even higher intensities envisaged at upcoming multi-petawatt class facilities \cite{Danson:2019}.
The interaction of an ultra-relativistic electron beam with electromagnetic fields of this magnitude represents an ideal experimental configuration \cite{Bula:1996, Burke:1997, Sarri:2014, Poder:2018, Cole:2018, Yan:2017, Abramowicz:2021, Salgado:2021, Turner:2022} to access unexplored regimes of strong-field quantum electrodynamics (SFQED), where the transition from perturbative to non-perturbative processes, as well as the transition from classical to quantum dynamics, occurs \cite{DiPiazza:2012, Blackburn:2020rr, gonoskov.rmp.2022, Fedotov:2023}. 

The non-linear nature of this regime is parameterized by the peak normalised field amplitude $a_0$, which is related to the electron quiver momentum, $\abs{\vb{p}_\perp}$, in the laser field as $a_0 = \abs{\vb{p}_\perp}/m$.
The relative importance of quantum contributions to the electron dynamics is determined by the quantum parameter, $\chi = e\abs{F_{\mu\nu}p^\nu}/m^3$. Here, $F_{\mu\nu} = \partial_\mu A_\nu - \partial_\nu A_\mu$ is the electromagnetic field tensor with potential $A_\mu$, $p^\nu$ is the electron momentum, and we assume natural units whereby $\hbar = c = 1$. Generally speaking, the SFQED regime is reached whenever $a_0\gg1$ and $\chi\gtrsim1$.
In the case of a head-on collision of an electron with energy $E$ with a plane wave of peak intensity $I_0$ and wavelength $\lambda$, the parameters $a_0$ and $\chi$ can be expressed, in the laboratory frame, as:  $a_0 = 6\lambda\,\bqty{\unit{\um}}\, \sqrt{I_0\,\bqty{10^{20}\ \unit{W/cm^2}}}$ and $\chi = (5.9\times10^{-2})\,E\,\bqty{\unit{\GeV}}\,\sqrt{I_0\,\bqty{10^{20}\ \unit{W/cm^2}}}$. 

Remarkably, formulation and calculation of a first-principle and accurate theory for the dynamics of an electron in an external electromagnetic field of arbitrary intensity is still one of the most fundamental outstanding problems in electrodynamics, with only sparse experimental studies reported to date at ultra-high intensities \cite{Poder:2018, Cole:2018, Yan:2017}.  SFQED phenomena (including non-linear Inverse Compton Scattering (ICS) \cite{Bula:1996,Sarri:2014,Yan:2017,Acosta:2021}, quantum radiation reaction \cite{Poder:2018,Cole:2018}, and Breit-Wheeler pair production \cite{Burke:1997}) are strongly dependent on laser intensity, which thus must be reliably determined to allow for a meaningful comparison between experimental results and different, sometimes competing, theoretical models.

The intensity of a focussed high-power laser can be indirectly inferred from separate measurements of its energy, pulse duration, and focal spot size \cite{Yanovsky:2008}, with the latter often measured only at a reduced power. 
This procedure is prone to significant uncertainties, and neglects some potentially important factors such as the presence of longitudinal fields in the tight focus of the laser \cite{Salamin2007} and possible electron-laser spatio-temporal misalignment. 
Alternative approaches involving photo-ionization of low-density gases \cite{LHuillier:1983,Hetzheim:2009} or measurement of the spatial profile of Thomson-scattered electrons \cite{Gao:2006, Ravichandran:2023} have also been proposed to determine the peak laser intensity, but are of limited applicability for the on-shot measurement of laser intensity in a SFQED experiment. 

Recently, it has been proposed that the characterization of the transverse angular profile of Compton-scattered photons could provide an estimation of the laser intensity at the interaction point \cite{HarShemesh:2012, Yan:2017, Harvey:2018}. The disadvantage of approaches of this kind is that they  depend on a specific SFQED model of the radiation reaction experienced by the electrons. As the appropriateness of the model is in turn determined by $a_0$, these approaches are of limited applicability when the underlying physics itself is under investigation, for instance when probing the transition regimes. 
A model-independent technique, where the laser intensity could be inferred from the energy-weighted angular profile of the photons, was proposed in \cite{Blackburn:2020} and it was shown to be able to retrieve a value for $a_0$ at the interaction point with an uncertainty of the order of a few percent. 
However, measuring the spatial distribution of the photon beam energy is experimentally challenging: detection of high-energy (i.e., MeV-GeV) photon beams is typically done at either high-flux and low energy (up to $\sim 1$ \unit{\MeV})~\cite{Wilhelm:1996, Lipoglavek:2006} or at high energies on a single photon basis~\cite{Ypsilantis:1994, Schonfelder:2013}.
This is mainly due to the typically weak dependence of detector response on photon energy in the multi-MeV range. Techniques to determine the energy spectrum of a high energy, high flux gamma-ray beam have also been proposed \cite{Schumaker:2014, Corvan:2014, Behm:2018,Cavanagh:2023}; however spatial information cannot be recovered.
This experimental limitation constrains the applicability of the method proposed in \cite{Blackburn:2020}, which relies on the energy density profile.
We thus propose here an inference method which relies instead on a precise knowledge of the transverse distribution of the photon number density, which can be readily obtained using existing technology.

By first considering the laser as a plane wave with a Gaussian temporal envelope, analytical expressions for the number-weighted angular distribution of Compton photons are derived, including corrections due to radiation reaction, resulting in a theoretical description which is model-independent over a large parameter space. Realistic effects such as the finite electron beam size, laser focusing, and transverse offsets are then included using geometric considerations of the interaction.
For well-characterized electron beams, this inference method is shown to provide an accurate single-shot estimate on the interaction $a_0$ for each collision. We show that the derived expressions closely match the results of numerical simulations, both in the case of plane wave and focused fields.
An uncertainty in retrieving $a_0$ of less than $15\%$ across the investigated range is demonstrated for the model, even with the inclusion of pair production, providing an ideal tool to infer the laser intensity at the interaction point on a shot-to-shot basis.

\section{Analytical Results}\label{sec:AnalyticalResults}
\begin{figure}[t]
    \centering
    \vspace{2.5mm}
    \includegraphics[width=\linewidth]{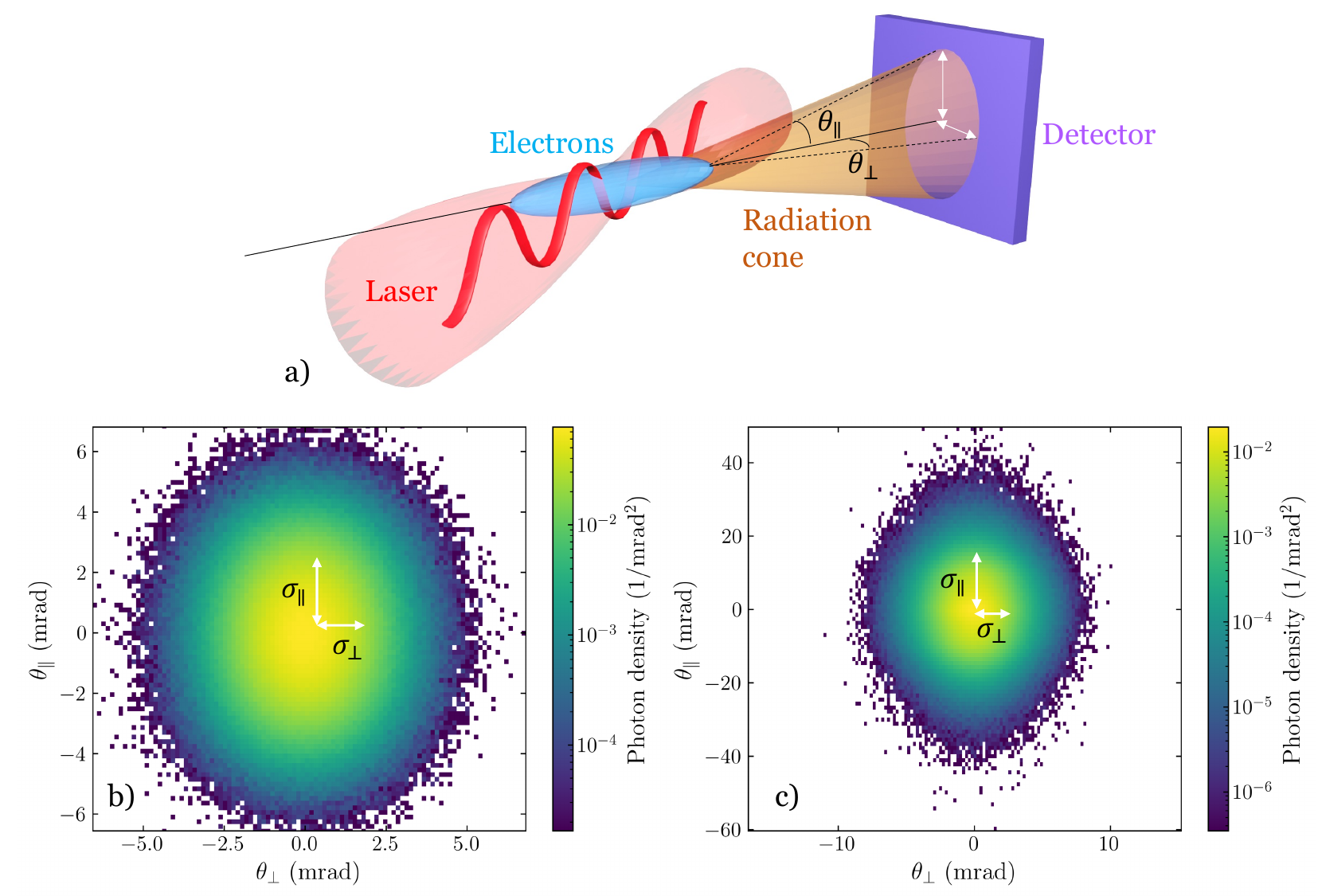}
    \caption{(a) A visualization of an electron bunch interacting with a linearly polarised laser as described in the text. Here, the laser is vertically polarised, with the electron bunch propagating in the horizontal plane. The resultant radiation cone and its transverse profile on a detector are shown, with the parallel and perpendicular emission angles, $\theta_\parallel$ and $\theta_\perp$, marked. Examples of the simulated angular distributions of the Compton-scattered photons for an electron energy of 1 GeV and a laser intensity of (b) $a_0 = 5$, and (c) $a_0 = 20$ are shown, with the direction of $\sigma_\parallel$ and $\sigma_\perp$ highlighted.}
    \label{fig:interaction_diagram}
    \vspace{2.5mm}
\end{figure}

We consider here the case of an electron (of charge $-e$, mass $m$, and energy $\gamma m$) colliding head-on with a monochromatic, linearly polarized electromagnetic wave of peak normalised amplitude $a_0$ and frequency $\omega_0$, where $\gamma \gg a_0 > 1$. 
Defining the laser phase as $\varphi$, the normalised potential of such a plane wave is given by $a_\mu = a_0 g(\varphi)\varepsilon_\mu \sin(\varphi)$, where $g(\varphi)$ is the pulse envelope and $\varepsilon_\mu$ is the polarization vector. The angle between the electron's momentum and the laser propagation axis is then given over a single cycle of the wave as $\theta(\varphi) = a_0\sin(\varphi)/\gamma$, lying in the plane containing the laser polarization (electric field) vector. 
Additionally, the electron quantum parameter for a counterpropagating plane wave can be expressed as $\chi(\varphi) = 2\gamma a_0 \omega_0 \vert\cos(\varphi)\lvert/m$. 
The photon emission rate, derived under the locally constant field approximation (LCFA), is proportional to the instantaneous electric-field strength, $\dot{N}_\gamma \propto a_0 \abs{\cos\varphi}$, with the dot representing differentiation with respect to laboratory time. 
As emission occurs primarily along the direction of the electron's momentum, the mean-square emission angle over a single cycle can be defined as
\begin{equation}
    \sigma_\parallel^2 \equiv \frac{\int\, \theta^2(\varphi) \dd{\dot{N}_\gamma}}{\int\, \dd{\dot{N}_\gamma}} 
    = \frac{a_0^2}{3\gamma^2} + \sigma_\perp^2.
\label{eq:MSqEmissionAngle}
\end{equation}
The subscripts $\parallel$  and $\perp$ refer to the directions parallel and perpendicular to the laser polarization, respectively (see \cref{fig:interaction_diagram}). The additional $\sigma_\perp^2$ term in \cref{eq:MSqEmissionAngle} is thus the mean square emission angle perpendicular to the polarisation plane as shown in \cref{fig:interaction_diagram}.
This accounts for the angular broadening induced by the electron beam divergence and for the fact that photons are not emitted exactly parallel to the electron's instantaneous velocity~\cite{blackburn.pra.2020}.

It should be noted that the reported expression for the photon  emission rate ($\dot{N}_\gamma \propto a_0\lvert\cos\varphi\lvert$), is strictly only valid for $\chi \ll 1$; for $\chi\gg1$, the photon emission rate scales as $\dot{N}_\gamma \propto \alpha \chi^{2/3}$. However, due to the normalising factor in the denominator of \cref{eq:MSqEmissionAngle}, this change of scaling does not affect the resulting form of the inference equation, $\sigma_\parallel^2 \sim a_0^2/\gamma^2$.

To include the effects of a temporal pulse envelope $g(\varphi)$ and of radiation reaction, the linearity of variances under addition can be used to sum the contribution to emission from each wave cycle. The contribution to the variance of the angular profile (in the plane containing the laser electric field) at a given phase is thus 
$\sigma_\parallel^2(\varphi) = a_0^2 g^2(\varphi) / \bqty{3 \gamma^2(\varphi)} + \sigma_\perp^2(\varphi)$.
Hence, by averaging this over the pulse profile and weighing by the photon emission rate, the total difference in the transverse variances is
\begin{equation}
    \sigma_\parallel^2 - \sigma_\perp^2 = \left.
        \int_{-\infty}^{\infty} \!
            \frac{a_0^2 g^2(\varphi)}{3 \gamma^2(\varphi)} \dv{N_\gamma}{\varphi}
        \, \dd{\varphi}
        \middle/
        \int_{-\infty}^{\infty} \!
            \dv{N_\gamma}{\varphi}
        \, \dd{\varphi}
        \right..
\end{equation}

To evaluate $\gamma(\varphi)$, we will first assume a classical radiation reaction model described by the Landau-Lifshitz equation \cite{Landau:1975}. In this case, the electron Lorentz factor evolves as \cite{DiPiazza:2008}
    \begin{align}
    \gamma(\varphi) &= \frac{\gamma_i}{1 + R_c \mathcal{I}(\varphi) / 3},
    &
    R_c &= \frac{2 \alpha a_0^2 \gamma_i \omega_0}{m},
    &
    \mathcal{I}(\varphi) &= \int_{-\infty}^\varphi g^2(\phi) \, \dd{\phi},
    \label{eq:ClassicalRR}
    \end{align}
where $\gamma_i$ is the initial Lorentz factor of the electron, and $\alpha$ is the fine structure constant. Thus
    \begin{equation}
    \sigma_\parallel^2 - \sigma_\perp^2 =
        \frac{a_0^2}{3 \gamma_i^2}
        \pqty{ \int_{-\infty}^{\infty} \! g(\varphi) \,\dd{\varphi} }^{-1}
        \int_{-\infty}^{\infty} \!
            g^3(\varphi) \bqty{ 1 + \frac{R_c}{3} \mathcal{I}(\varphi) }^2
        \, \dd{\varphi}.
    \label{eq:ClasVarDiffDef}
    \end{equation}
In general, the second integral cannot be solved analytically, however for flat-top and Gaussian pulses, the following closed-form solution is found:
    \begin{equation}
    \sigma_\parallel^2 - \sigma_\perp^2 =
        \frac{a_0^2}{3 \kappa_1}
        \bqty{
            \frac{1}{\gamma_i \gamma_f}
            + \kappa_2 \pqty{ \frac{1}{\gamma_f} - \frac{1}{\gamma_i} }^2
        },
    \label{eq:ClasVarDiff}
    \end{equation}
with $\kappa_1 = 1$ ($\sqrt{3}$) and $\kappa_2 = 1/3$ ($0.315$) for flat-top (Gaussian) pulses respectively, and $\gamma_f$ is the final Lorentz factor of the electron after passing through the field. 

\begin{figure}[b]
    \centering
    \vspace{2.5mm}
    \includegraphics[width=0.65\linewidth]{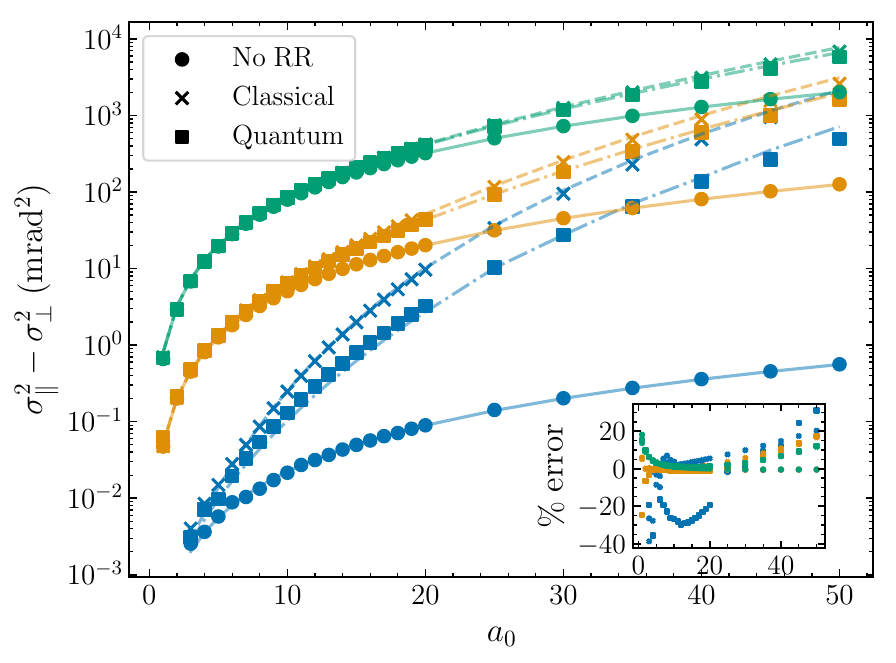}
    \caption{Difference in the variance parallel and perpendicular to the laser polarisation axis of the emitted radiation profile by an electron beam with central energy, $\gamma_i m$, $1\%$ RMS energy spread and divergence $\delta = 1\ \unit{\mrad}$ for: $\gamma_i m = 250$ \unit{\MeV} (green); $\gamma_i m=1$ \unit{\GeV} (orange); and $\gamma_i m=15$ \unit{\GeV} (blue) as predicted by Eq. \ref{eq:QedVarDiff} (lines) and calculated from LCFA simulations (points). Different radiation reaction models are considered: no (dots), classical (crosses), and quantum (squares) radiation reaction. The laser is modelled as a plane wave with a Gaussian envelope with FWHM duration, $\tau = 40$ \unit{\fs} and the threshold in emitted photon energy was $\omega_{\text{min}}'=1$ \unit{\MeV}. The inset shows the relative difference between the numerical and analytical results.}
    \label{fig:profile_variance_difference}
    \vspace{2.5mm}
\end{figure}

Quantum effects affect radiation reaction mainly in two ways \cite{Vranic:2016, Ridgers:2017}: first, the average energy loss by radiation is reduced by the Gaunt correction factor \cite{Kirk:2009}; and second, the stochastic nature of emission means that $\gamma_f$ is no longer a single-valued function of $\gamma_i$.
For electrons with an initial energy spectrum $dN_e/d\gamma_i$, weighting \cref{eq:ClasVarDiff} by $(\gamma_i - \gamma_f)\frac{dN_e}{d\gamma_i}$ and integrating over all $\gamma_i$ accounts for the increased emission power by electrons with higher instantaneous $\gamma$.
Then, under the same assumptions as \cite{Blackburn:2020}, we arrive at an equivalent of \cref{eq:ClasVarDiff} that takes into account quantum radiation reaction and a finite initial energy spread:
\begin{align}
\begin{split}
\sigma_\parallel^2 - \sigma_\perp^2 &=
    \frac{a_0^2}{3 \kappa_1}
    \bqty{
        \expval{\frac{1}{\gamma_i}} \expval{\frac{1}{\gamma_f}}
        + \kappa_2 \Bqty{ \expval{\frac{1}{\gamma_f^2}} + \expval{\frac{1}{\gamma_i^2}} - 2\expval{\frac{1}{\gamma_i}} \expval{\frac{1}{\gamma_f}} }
    } \\
    &\equiv \frac{a_0^2}{3\kappa_1}\mathcal{F}(\gamma_i, \gamma_f; \kappa_2).
\label{eq:QedVarDiff}
\end{split}
\end{align}
The moments over the final Lorentz factor $\gamma_f$ distribution appearing in this expression account for the stochastic broadening of the energy spectrum and are combined into a single function that depends only on the electron dynamics, $\mathcal{F}(\gamma_i, \gamma_f; \kappa_2)$. We also neglect the formation of electron-positron pairs via the non-linear Breit-Wheeler process by the scattered photons. The consequences of this, particularly in the $\chi \gtrsim 1$ regime, are discussed in \cref{sec:Discussion}.

In order to assess the accuracy of these analytical results, we compare the angular variances predicted by \cref{eq:QedVarDiff} with the results of numerical simulations in \cref{fig:profile_variance_difference}.
Using the Monte-Carlo code \textsc{Ptarmigan} \cite{Ptarmigan, Blackburn:2023}, we simulated the head-on  interaction of an electron beam with a plane-wave laser pulse with a Gaussian temporal envelope of full-width at half maximum (FWHM) duration $\tau = 40$ \unit{fs} and wavelength $\lambda = 0.8$ \unit{\um}. 
A cylindrically symmetric electron beam with a radius of $r_b = 0.5$ \unit{\um}, an RMS divergence of $1\ \unit{\mrad}$, a $1\%$ RMS energy spread, and different mean energies (250 \unit{\MeV}, 1 \unit{\GeV} and 15 \unit{\GeV} in green, yellow and blue respectively) was considered. 
Additional simulations with an initial energy spread up to $\sim 10\%$ showed no appreciable effect on the results.
Three models of radiation reaction were used in the simulations: no radiation reaction, so that the particle trajectory is governed by the Lorentz force; classical radiation reaction using the Landau-Lifshitz equation; and a fully stochastic, quantum model using emission rates calculated in the LCFA \cite{Ritus:1985}. 

As low energy photons can be emitted at large angles, an energy cut is introduced when extracting $\sigma_\parallel^2 - \sigma_\perp^2$ from the simulations, in order to remove the divergent behaviour of the distribution variance, particularly for $a_0 \sim \mathcal{O}(1)$. 
This threshold energy can be expressed as $\omega_\text{min}' = f_\text{min}\gamma(\varphi)m$, where $0 < f_\text{min} < 1$. 
For this work, we used a nominal value of $\omega_\text{min}' = 1\, \unit{\MeV}$. 
However, the results reported here are robust over a wide range of low-energy cut-offs with higher order corrections in $f_{min}$ showing to have a negligible effect (see  \cref{sec:SpectralCutOff}). The precise implementation of such an energy cut in practice is experiment-specific. 
As an example, \cite{Abramowicz:2023} describes a setup with a thin (500 \unit{\um}) aluminium window and reports that photons with an energy less than $\sim$ 1 \unit{\MeV} are sufficiently absorbed or scattered out of the photon cone to be profiled.
The window has a negligible scattering effect on the high energy ($\gtrsim 100\ \unit{\MeV}$) photons.

Good agreement between the analytical prediction of \cref{eq:QedVarDiff} and the simulation results can be seen across a range of six orders of magnitude for each radiation reaction model and initial electron energy
with a relative error between \cref{eq:QedVarDiff} and the numerical results of less than $\sim$25\%.
The simulated results include a lower energy threshold at 1 \unit{\MeV}, indicating that applying an energy cut much less than the electron energy has a negligible effect on the calculation of $\sigma_\parallel^2 - \sigma_\perp^2$.  
For electron energies $\lesssim 1$ \unit{\GeV}, the classical and quantum models give similar results, however for larger electron energies (and hence larger $\chi$), the classical model overpredicts the broadening compared to the quantum model. 
This is due to the ``Gaunt-modified'' emission spectrum of the quantum model compared to the classical model.
Additionally, for $\chi \gtrsim 0.1$, stochastic effects become non-negligible and radiation is no longer continuous as in the classical model \cite{Neitz:2013,Vranic:2016,Ridgers:2017}.

Calculation of $\sigma_\parallel^2 - \sigma_\perp^2$ requires knowledge of the initial and final distributions of the electron energy to be able to determine the appropriate moments, which is straightforward for simulation data.
Experimental measurement of the final energy spectrum is also uncomplicated, even on a shot-to-shot basis; however, non-invasive measurement of the initial electron spectrum is a non-trivial task. 
Techniques such as those outlined in \cite{Baird:2019, Emma:2021,Streeter:2023} may be used in cases where the beam properties are expected to significantly vary shot-to-shot, such as in laser-wakefield acceleration (LWFA) experiments.
Alternatively, the distribution can be measured in the absence of the laser pulse, for sufficiently stable electron beams.

\section{Intensity Inference}
\subsection{Plane Waves}\label{sec:PlaneWaves}
By measuring the angular size of the radiation profile, it is thus possible to infer the laser intensity, i.e. the interaction $a_0$, using \cref{eq:QedVarDiff}. 
This inference depends on the mean initial and final electron energies, with any explicit radiation reaction dependence being absorbed into the latter. 
Rearrangement of \cref{eq:QedVarDiff} gives
\begin{equation}
    a_0^2 =
        \frac{3 \kappa_1}{\mathcal{F}(\gamma_i,\gamma_f; \kappa_2)}
(\sigma_\parallel^2 - \sigma_\perp^2).
\label{eq:IntensityInfer}
\end{equation}
Since the intensity inference depends on the difference in the parallel and perpendicular variances, the effect of electron divergence $\delta$ is removed as this is contained in both $\sigma_\parallel^2$ and $\sigma_\perp^2$, under the assumption of a cylindrically symmetric electron beam. The divergence of the electron beam will also have an indirect effect on the overlap of the electron beam and laser focal spot, which is discussed in more detail in \cref{sec:FocusedFields}.

\begin{figure}[b]
    \centering
    \vspace{2.5mm}
    \includegraphics[width=0.65\linewidth]{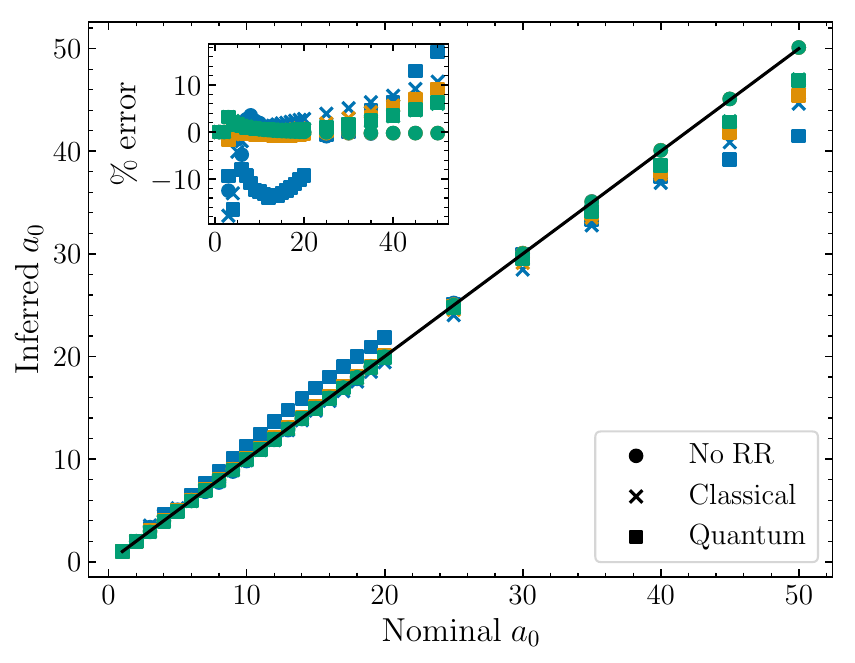}
    \caption{Inferred $a_0$, calculated from the simulation results using \cref{eq:IntensityInfer},  versus nominal $a_0$ (and percentage error, inset) for different radiation reaction models. The simulations assume an electron beam with a $\delta = 1\ \unit{\mrad}$ divergence, a $1\%$ energy spread, and different mean energies: $\gamma_i m = 250$ \unit{\MeV} (green); $\gamma_i m = 1$ \unit{\GeV} (orange); and $\gamma_i m = 15$ \unit{\GeV} (blue). Laser pulse was modelled as a plane wave with a Gaussian envelope of FWHM duration, $\tau = 40$ \unit{\fs}. The black solid line is plotted as a guide for the eye. The inset shows the relative difference between the numerical and analytical results.}
    \label{fig:intensity_inference}
    \vspace{2.5mm}
\end{figure}

\cref{fig:intensity_inference} shows the result of applying \cref{eq:IntensityInfer} to the same simulated electron-laser configuration as in the previous section for the different electron energies and radiation reaction models. 
Across the range under investigation, the inference works remarkably well, within $10\%$ across the entire range of intensities and each radiation reaction model, with the only exception of $a_0 < 5$ and high electron energy.
As expected, the no radiation reaction model is the most precise in this case, estimating the peak intensity to $\sim 1\%$ accuracy.
For the classical and quantum models, \cref{eq:IntensityInfer} begins to underestimate the correct value as $a_0$ increases. 
Moreover, for large $a_0$, the radiative cooling of the electrons may result in the assumption $\gamma \gg a_0$ no longer holding. 
At this point, the inference becomes less accurate as the typical emission energy is of similar magnitude to the employed energy cut.
This is discussed further in \cref{sec:Discussion}. Simulations also show that detecting photons within the angular range $\abs{\theta} \lesssim 2a_0/\gamma_f$ is necessary for accurate inference. 

The uncertainty in inferring the intensity using \cref{eq:IntensityInfer} due to measurement of the electron energies and of the profile variances can be expressed as:
\begin{equation}\label{eq:InferUncertainty}
\begin{gathered}
    \pqty{\frac{\delta a_0}{a_0}}^2 = \frac{(\sigma_\parallel^2 + \sigma_\perp^2)\delta\sigma^2}{(\sigma_\parallel^2 - \sigma_\perp^2)^2} + \frac{1}{4}\pqty{\frac{\delta\mathcal{F}}{\mathcal{F}}}^2, \\
    \frac{\delta\mathcal{F}}{\mathcal{F}} = \frac{(1-2\kappa_2)^2 A + \kappa_2^2 B}{C},\\
    A = \Gamma_{i,1}^2\delta\Gamma_{i,1}^2 + \Gamma_{f,1}^2\delta\Gamma_{f,1}^2,\qquad B = \delta\Gamma_{i,2}^2 + \delta\Gamma_{f,2}^2, \\
    C = \bqty{\Gamma_{i,1}\Gamma_{f,1}+3\kappa_2(\Gamma_{i,2}+\Gamma_{f,2} - 2\Gamma_{i,1}\Gamma_{f,1})}^2,
\end{gathered}
\end{equation}
where $\Gamma_{\ell, n} = \expval{\gamma_\ell^{-n}}$ are the negative moments of the initial and final electron gamma factors. 
For $a_0 \sim \mathcal{O}(1)$, the dominant source of uncertainty comes from the measurement of the transverse variances. 
In particular for this intensity range, $(\sigma_\parallel^2 - \sigma_\perp^2)^2$ can become very small, resulting in a large uncertainty on the inferred $a_0$ due to the negligible ellipticity of the gamma-ray beam. 
For $a_0 \gtrsim 10$, the ellipticity of the radiation profile increases with $a_0$, and so the uncertainty in inferring $a_0$ due to the measurement of $\sigma_\parallel$ and $\ \sigma_\perp$ decreases approximately quadratically.

\subsection{Focused Fields}\label{sec:FocusedFields}
Intensity measurements in a more realistic scenario require consideration of the spatial and temporal structure of a focused laser, as well as the finite size of the electron beam. 
For instance, if the laser is focused to a spot of $1/e^2$ radius $w_0$ comparable to or smaller than the transverse RMS radius $r_b$ of the electron beam, the electrons will experience a distribution of intensities. In this case, the inferred $a_0$ will be reduced compared to the true peak $a_0$ as the laser intensity distribution will be effectively averaged over the spatial profile of the electron beam.
To account for this geometrical effect, we consider a cylindrically symmetric electron beam with a Gaussian transverse profile of radius $r_b$ and transverse offset $x_b$ compared to the centroid of the laser focal spot. 
The field intensity experienced by a single electron at a transverse position $(x,y)$ compared to the laser peak intensity will be $a(x,y) = a_0 \exp[-(x^2 + y^2)/w_0^2)]$.
Averaging over the profile of the electron beam, the inferred intensity will be given by the average, to lowest order in $\alpha$, $a_0^\mathrm{inf} = \sqrt{\expval{a^3} / \expval{a}}$.
This inferred $a_0$ is then,
\begin{equation}
    a_0^\mathrm{inf} = a_0\sqrt{\frac{P}{Q}}\exp\pqty{ -\frac{\zeta^2}{PQ} },
\label{eq:CorrectedInference}
\end{equation}
where, $P = 1 + 3\rho^2$, $Q = 1 + 6\rho^2$ and $\rho = r_b/w_0$, $\zeta = x_b/w_0$.
This geometrical correction \cref{eq:CorrectedInference} is similar to that found in \cite{Blackburn:2020}, with differences only in the numerical factors due to the different scalings in photon number and power emission rates (proportional to $a_0$ and $a_0^2$, respectively). 

\begin{figure}[t]
    \centering
    \vspace{2.5mm}
    \includegraphics[width=\linewidth]{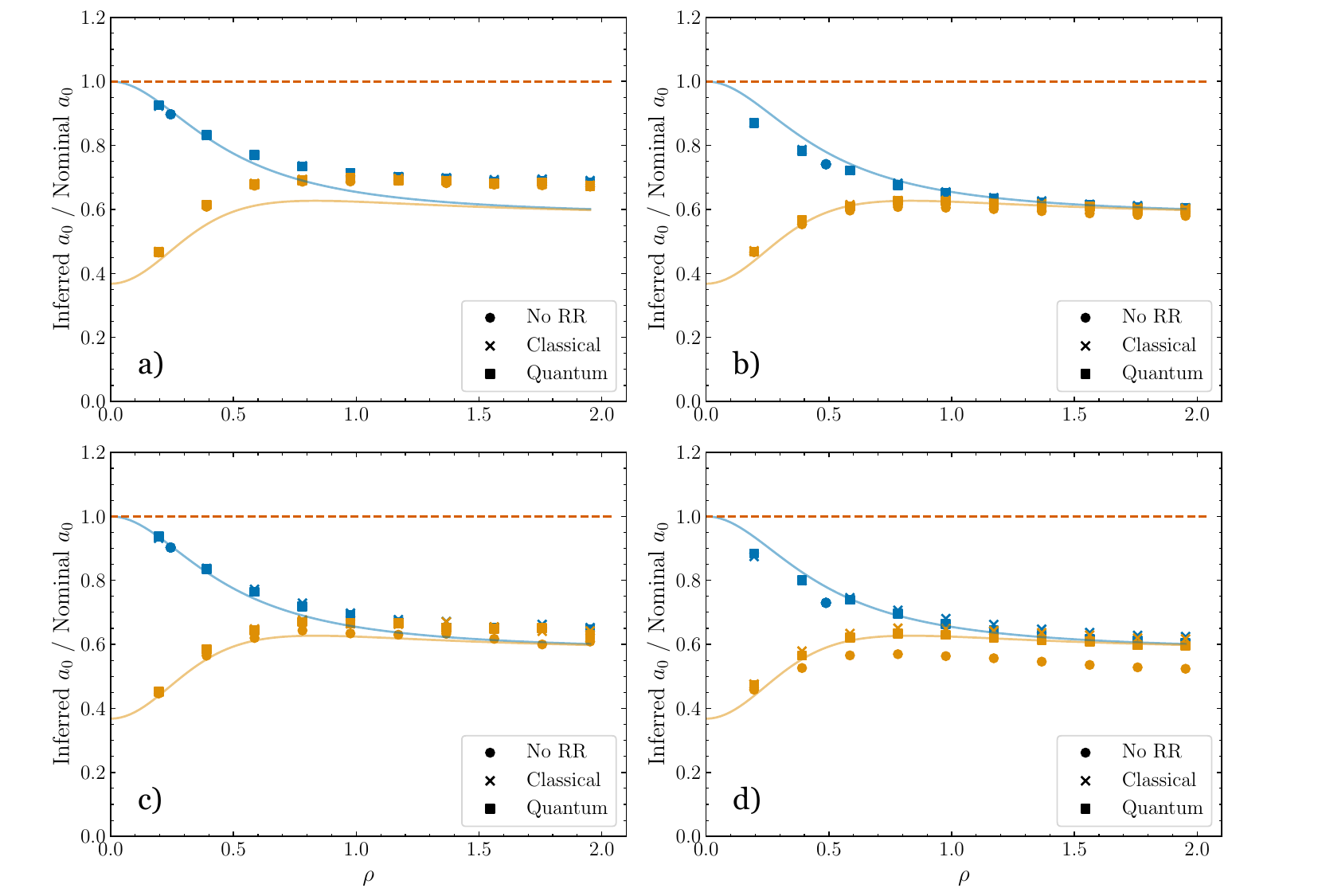}
    \caption{Fraction of inferred $a_0$ to true value as a function of increasing electron beam size for $x_b = 0$ (blue) and $x_b = w_0$ (yellow). Electron beam had a mean energy of: $250$ \unit{\MeV} (a, b) and $1$ \unit{GeV} (c,d) with 1\% RMS spread. The laser pulse was modelled with a Gaussian spatiotemporal profile with waist $w_0 = 2.04$ \unit{\um} (a,c) and $1.02$ \unit{\um} (b,d) and FWHM duration $\tau = 40$ \unit{\fs}, corresponding to a peak intensity of $a_0 = 10$ and $20$, respectively. Different radiation reaction models were also considered. The dashed red line is plotted as a guide for the eye.} 
\label{fig:focused_field_inference}
    \vspace{2.5mm}
\end{figure}

To determine if this reduction in the inferred intensity is well accounted by \cref{eq:CorrectedInference}, we performed a series of simulations of a 250 \unit{\MeV} and a 1 \unit{\GeV} electron beam with a transverse Gaussian profile of radius, $r_b$, colliding head-on with a laser pulse having a Gaussian intensity profile focused to a spot size $w_0$. 
As examples, we show simulation results in \cref{fig:focused_field_inference} as a function of the electron beam radius, assuming two laser intensities ($a_0 = 10$ and $a_0 = 20$), two transverse offsets ($x_b = 0$ and $x_b = w_0$), and different radiation reaction models. 
Increasing the electron beam size has an observable effect on the inference, resulting in a reduction by $\sim 30\%$ for $\rho \gtrsim 1$. This is consistent with the limit of \cref{eq:CorrectedInference} to be $1/\sqrt{2}$ for $\rho\rightarrow \infty$. Conversely, for misaligned beams, increasing $r_b$ improves the inferred $a_0$ value, as progressively higher intensity regions in the laser field distribution will be experienced by the electrons. For large $\rho$, more of the electron beam interacts with the low intensity fringes of the laser pulse, where the local $\chi$ will be low, resulting in low energy, high angle photons. 
These low energy photons are then removed by the energy threshold, resulting in a smaller angular profile and hence an underestimate in the inference of $a_0$. 
The increased ponderomotive scattering due to radiation reaction acts to compensate with this underestimation in both the classical and quantum models.
Similar simulations for different electron energies and peak $a_0$ (not shown) yield similar conclusions.

If the peak laser intensity is measured by an alternative means, such as those mentioned in \cref{sec:Introduction}, any misalignment in the form of a transverse spatial offset could be estimated from the reduction in the inferred intensity $a_0^\mathrm{inf}$ with respect to the nominal $a_0$. 
Similarly to \cite{Blackburn:2020}, producing a distribution of $a_0^{\mathrm{inf}}$ on different shots is sufficient to monitor imperfect pointing stability, and any systematic effects such as finite electron beam size can be deduced.

\section{Discussion}\label{sec:Discussion}
While the nominal $a_0$ can be inferred to within $10\%$ for most of the parameter space explored here, there are a few caveats to the proposed approach. As \cref{eq:ClasVarDiff} was derived under the assumption of LCFA rates, the use of \cref{eq:IntensityInfer} in the regime $a_0 \sim \mathcal{O}(1)$ is anticipated to give less accurate results \cite{Blackburn:2021, Fedotov:2023}; the LCFA overpredicts the emission of low energy - and, hence, higher divergence - photons in this range resulting in an overestimation in the inferred $a_0$.
A correction for $a_0 \sim \mathcal{O}(1)$ using the more accurate locally monochromatic approximation (LMA) \cite{Bamber:1999,Chen:1995,Heinzl:2020,Torgrimsson:2021,King:2021} is the subject of current work. 

\begin{figure}[tb]
    \centering
    \vspace{2.5mm}
    \includegraphics[width=0.65\linewidth]{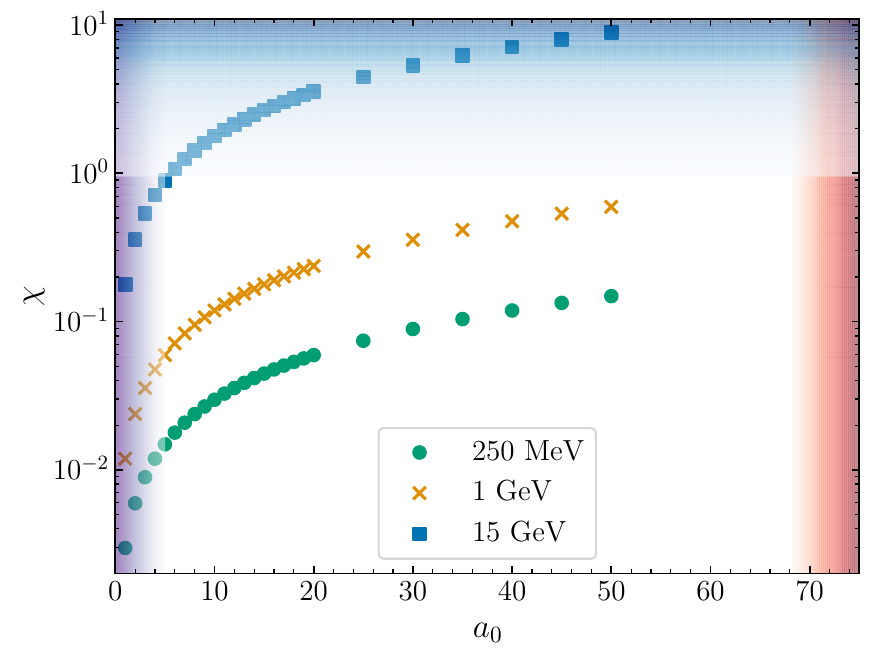}
    \caption{Plot showing the regions of $a_0 - \chi$ parameter space where the assumptions used to derive \cref{eq:IntensityInfer} are less accurate due to the use of LCFA (purple shaded region), neglecting pair production (blue shaded region), and introducing an energy cut-off in the photon detection (shaded red region). Simulated parameters are marked as points for mean electron energy 250 \unit{\MeV} (green), 1 \unit{\GeV} (orange) and 15 \unit{\GeV} (blue).}
    \label{fig:validity_range}
    \vspace{2.5mm}
\end{figure}

As $a_0$ increases, the electrons lose more energy via radiation which reduces the typical
photon emission energy over the interaction. If the selected energy cut $\omega_\text{min}'$ becomes comparable to the mean peak synchrotron energy \cite{Bell:2008}, the accuracy of the inference
method is reduced since a non-negligible fraction of the photon distribution would be removed. The range of $a_0$
which can be accurately inferred for a given energy cut can thus be estimated in the following way.
We can impose the (mean) peak synchrotron energy $\omega_\text{peak} = 0.44\gamma m \chi$ to be at least twice the energy cut. Using the definition of the electron quantum parameter, this requirement becomes $\gamma^2 a_0 \gtrsim 2.27\omega_\text{min}'/\omega_0$.
For a Gaussian envelope of FWHM duration $\tau$, radiation reaction can be accounted for by \cref{eq:ClassicalRR} with $\mathcal{I}(\infty) = \omega_0\tau\sqrt{\pi/\displaystyle 4\ln2}$, resulting in the valid intensity range
\begin{equation}\label{eq:valid_a0_range}
    a_0 \lesssim 860 \frac{\lambda\, [\unit{\um}]}{\tau^{2/3}\,[\unit{\fs}]\, \omega_\text{min}'^{2/3}\, [\unit{\MeV}]}.
\end{equation}
For a laser of wavelength $\lambda = 0.8\, \unit{\um}$ and FWHM duration $\tau = 40\, \unit{\fs}$, a 1 \unit{\MeV} energy cut will provide accurate inference of the intensity for $a_0 \lesssim 70$, as depicted in \cref{fig:validity_range}.
Decreasing the energy threshold to 0.01 \unit{\MeV} increases the range to $a_0 \lesssim 350$.
It should be noted that reducing the energy cut at low $a_0$ will tend to decrease the inference accuracy here.
 It is thus important to select an energy threshold appropriate for the intensity range under investigation and suitable for the specific experimental configuration.

\begin{figure}[tb]
    \centering
    \includegraphics[width=0.6\linewidth]{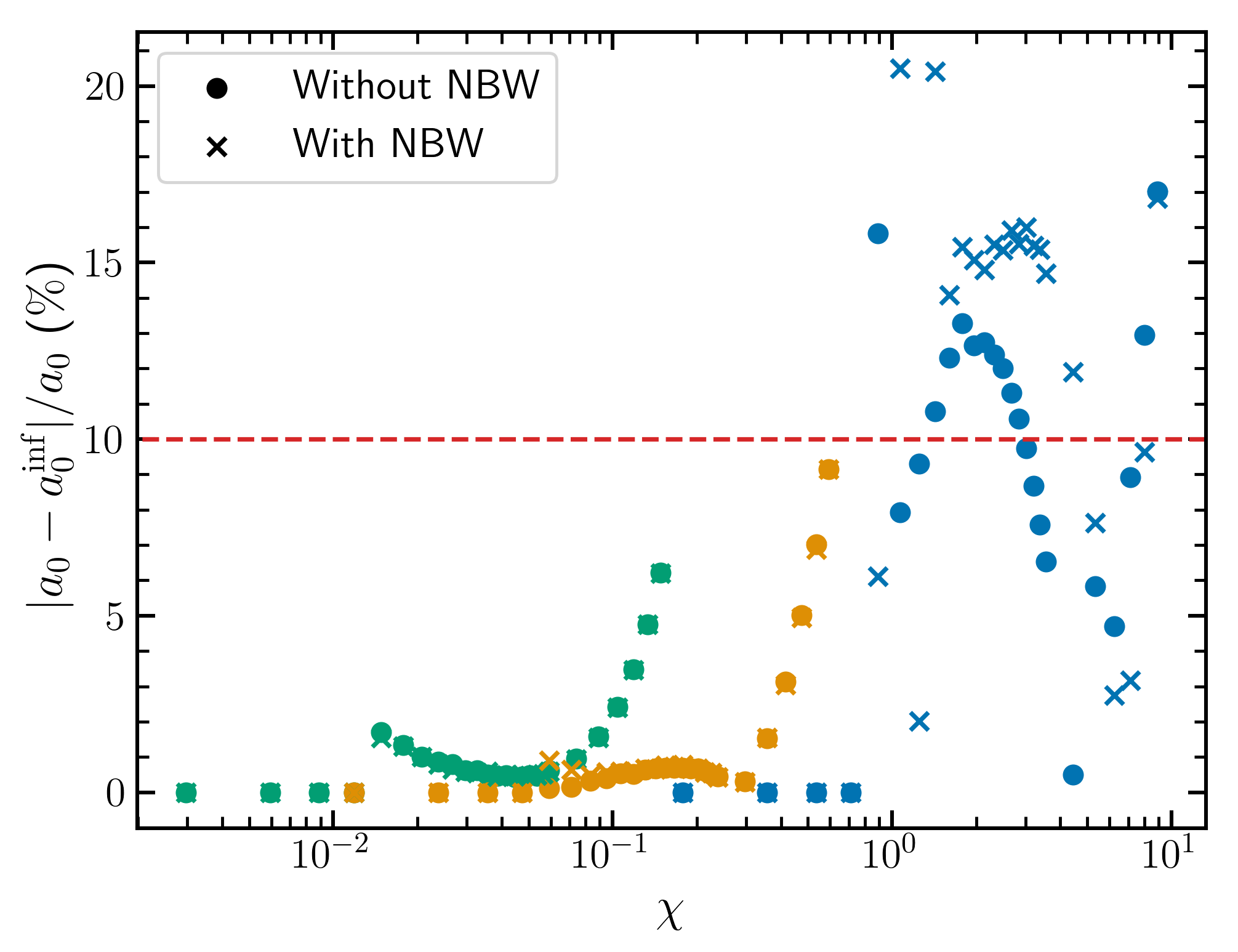}
    \vspace{2.5mm}
    \caption{Relative error in the value of inferred $a_0$ for the same parameters as \cref{fig:intensity_inference} with a quantum radiation reaction model. Simulations were performed for $\gamma_i m = 250\ \unit{\MeV}$ (green), $1\ \unit{\GeV}$ (orange), and $15\ \unit{\GeV}$ (blue), both with and without non-linear Breit-Wheeler (NBW) pair production enabled.}
    \label{fig:pair_effect}
    \vspace{2.5mm}
\end{figure}

 In the derivation of \cref{eq:QedVarDiff}, we neglected the effect of pair production on the profile of the photon beam.
 This is valid for $\chi \ll 1$ as production is exponentially suppressed in this regime.
 However, for $\chi \gtrsim 1$, electron-positron generation can become significant. 
 \cref{fig:pair_effect} shows how the relative error in inferring $a_0$ from the photon beam profile is affected by enabling pair production in \textsc{Ptarmigan}.
 For $\chi < 1$, the effect of pair production is negligible as expected.
 As $\chi$ increases above $\mathcal{O}(1)$, the inference becomes less accurate, both without and with pair production enabled.
 Overall, in this regime, using \cref{eq:IntensityInfer} results in a maximum relative error $\sim 15-20\%$.
 This indicates that, in this regime, it is difficult to distinguish the effects due to stochastic energy loss of the electron beam even without the presence of addition electron-positron pairs.
 This is also true on an analytical level as it has not yet been shown how to include the kinematic effect of pair generation on the transverse photon profile.

\section{Conclusions}\label{Conclusions}
In summary, we have shown that the number-weighted radiation profile produced by an ultrarelativistic electron beam traversing an intense counter-propagating laser pulse can be used to determine the peak laser intensity at interaction.
Additionally, we have shown that the inference can be applied with consistent results for different radiation reaction models; this is particularly advantageous for experiments which aim to test different dynamic models.
This inference technique has been analysed over a large range of intensities and electron energies, resulting in a tested range of the quantum parameter over 3 orders of magnitude ($\sim 10^{-2} - 10$). 

Recently, advancements have made Bayesian methods popular for handling measurements of extensive parameter spaces.
However, definitive observation of strong-field QED effects or radiation reaction at a high confidence level will rely on combining different measurements of the collision parameters as cross-signal consistency enhances the significance of the results.
Extraction of the collision $a_0$ via the method presented here is therefore a crucial component of this analysis.
Moreover, \cite{Olofsson:2023} present a novel approach which could be applied in upcoming experiments although there is a large computational cost with this framework, excluding the possibility of an on-shot implementation.
In particular, as the quantities in \cref{eq:IntensityInfer} can be measured independently, and on a single shot basis, this approach is ideal for diagnosing variations in $a_0$ in a shot-to-shot manner.

Furthermore, the model can be used to monitor, on a shot-to-shot basis, potential beam misalignment at the interaction point. 
We therefore present this inference method as a complementary tool to Bayesian approaches.

Additionally, radiation reaction is, as yet, still not well understood experimentally.
Given an alternative measurement of $a_0$, \cref{eq:QedVarDiff} could be used to provide insight on the effect of radiation reaction on the electron motion by considering the photon profile.

The datasets required to reproduce the analysis and figures presented here are available at \cite{dataset}.

\section{Acknowledgements}
G. S. wishes to acknowledge support from EPSRC (grant numbers EP/T021659/1 and EP/V049186/1). The authors are grateful for useful discussions within the LUXE and E-320 collaboration.

\bibliography{references}

\appendix
\section{Spectral Cut-Off}\label{sec:SpectralCutOff}
\Cref{eq:ClasVarDiff} is derived under the assumption that a non-zero $f_\text{min}$ has a negligible effect on the difference in the variances. The effect of $f_{min}$ can be estimated by considering the emission rate as a function of $\chi$ and $f_\text{min}$. Assuming that $f_\text{min} \ll \chi$,
    \begin{equation}
    \dv{N_\gamma}{t} = \frac{5 \alpha m \chi}{2\sqrt{3} \gamma}
        \bqty{
        1 - 1.076 \pqty{ \frac{f_\text{min}}{\chi} }^{1/3} + 0.2309 \, \frac{f_\text{min}}{\chi} + \cdots
        }.
    \label{eq:EmissionRate}
    \end{equation}
The emission rate falls to 50\% and then 10\% of its uncorrected value when $f_\text{min} / \chi = 0.118$ and $1.35$ respectively.
Note that for a pulse with a nontrivial envelope, $\chi \to 0$ as $\varphi \to \pm\infty$, whereas $f_\text{min}$ is always finite.
Thus in the fringes of the pulse, $f_\text{min} \not\ll \chi$ and the perturbative expansion is not accurate.
Then, the parallel variance at a given phase becomes
    \begin{equation}
    \sigma_\parallel^2(\varphi) = \frac{a_0^2 g^2(\varphi)}{3 \gamma^2(\varphi)}
        \Bqty{ 1 - 0.140 \bqty{ \frac{f_\text{min}(\varphi)}{\chi(\varphi)} }^{1/3} }
        + \sigma_\perp^2,
    \end{equation}
where $f_\text{min}(\varphi) = \omega'_\text{min} / [\gamma(\varphi) m]$ and $\chi(\varphi) = 2 a_0 \gamma(\varphi) g(\varphi) \omega_0 / m$.
Adding a low-energy cut-off reduces the expected variance, because low-energy photons are generally emitted at larger angles.
Averaging this over the pulse envelope, the following correction to \cref{eq:ClasVarDiff}, at lowest order in $f_\text{min} / \chi$, is obtained:
    \begin{align}
    \Delta \pqty{ \sigma_\parallel^2 - \sigma_\perp^2 } &\simeq
        \frac{a_0^2}{3 \gamma_i^2}
        \pqty{ \frac{f_\text{min}}{\chi} }^{1/3} 
        \frac{1.076 \, A B - 1.216 \, C D}{D^2},
    \\
    A &= \int_{-\infty}^{\infty} \!
            g^3(\varphi) \bqty{ 1 + \frac{R_c}{3} \mathcal{I}(\varphi) }^2
        \, \dd{\varphi},
    \\
    B &= \int_{-\infty}^{\infty} \! g^{2/3}(\varphi) \bqty{ 1 + \frac{R_c}{3} \mathcal{I}(\varphi) }^{2/3} \, \dd{\varphi},
    \\
    C &= \int_{-\infty}^{\infty} \! g^{8/3}(\varphi) \bqty{ 1 + \frac{R_c}{3} \mathcal{I}(\varphi) }^{8/3} \, \dd{\varphi},
    \\
    D &= \int_{-\infty}^{\infty} \! g(\varphi) \, \dd{\varphi},
    \label{eq:VarDiffCorrection}
    \end{align}
where $f_\text{min}$ and $\chi$ are now defined in terms of the initial electron energy and peak laser amplitude, i.e. $f_\text{min} = \omega'_\text{min} / (\gamma_i m)$ and $\chi = 2 a_0 \gamma_i \omega_0 / m$.
In the absence of radiation reaction, for a flat-top pulse:
\begin{equation}
    \lim_{R_c \to 0} \Delta \pqty{ \sigma_\parallel^2 - \sigma_\perp^2 } \simeq
        -0.151 \frac{a_0^2}{3 \gamma_i^2} \pqty{ \frac{f_\text{min}}{\chi} }^{1/3},
\end{equation}
and for a Gaussian pulse:
\begin{equation}
    \lim_{R_c \to 0}  \Delta \pqty{ \sigma_\parallel^2 - \sigma_\perp^2 } \simeq
        0.0162 \frac{a_0^2}{3 \gamma_i^2} \pqty{ \frac{f_\text{min}}{\chi} }^{1/3}.
\end{equation}
The correction is particularly small for the Gaussian pulse. If radiation reaction is included, this correction must be evaluated numerically from the expressions above.

\begin{figure}[b]
    \centering
    \vspace{2.5mm}
    \includegraphics[width=\linewidth]{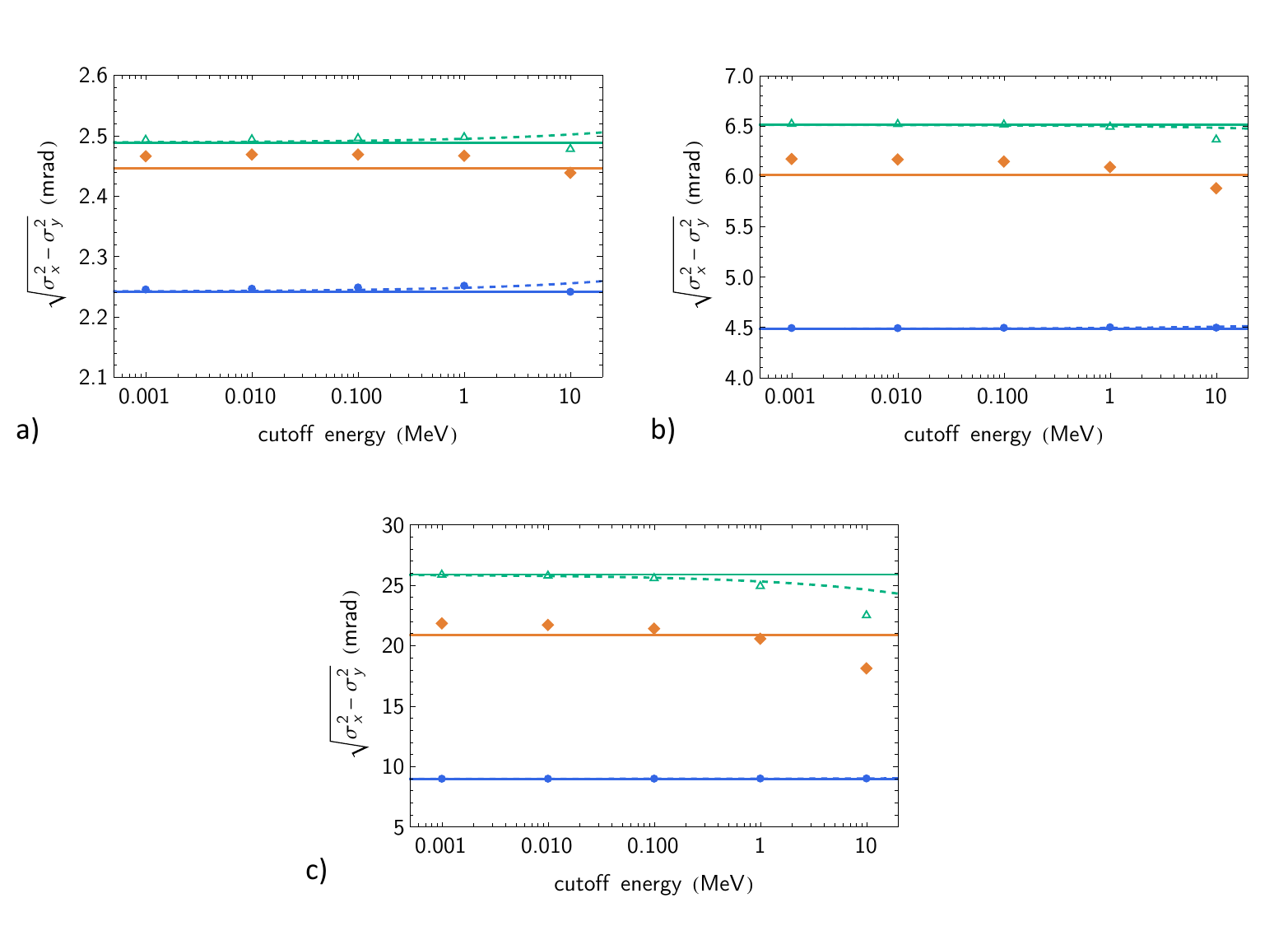}
    \caption{Results from theory (lines) and LCFA-based simulations (points) for 1 \unit{\GeV} electron beams colliding with 30 \unit{\fs} Gaussian pulses where (a) $a_0 = 10$, (b) $a_0 = 20$, and (c) $a_0=40$.
        Blue, green and orange indicate no radiation reaction, classical radiation reaction and quantum radiation reaction, respectively.
        The solid lines give the predictions of \cref{eq:ClasVarDiff} and \cref{eq:QedVarDiff}, as appropriate.
        The dashed lines include the correction from the low-energy cutoff, \cref{eq:VarDiffCorrection}, for the no-radiation reaction and classical radiation reaction cases.}
    \label{fig:SpectralCut}
    \vspace{2.5mm}
\end{figure}

In fact, \cref{fig:SpectralCut} shows that, for $a_0 \gtrsim 10$, corrections to the emission rate do not need to be considered for energy cuts less than the synchrotron peak energy, \cref{eq:EmissionRate}, as mentioned in \cref{sec:AnalyticalResults}.
For our choice of $\omega_\text{min}' = 1\, \unit{\MeV}$, the correction is of the order of a few percent, and so has a negligible effect on the results presented here.

\end{document}